\documentclass[preprint,showpacs,preprintnumbers,amsmath,amssymb]{revtex4}

\usepackage{graphicx}
\usepackage{dcolumn}
\usepackage{bm}

\begin{document}

\title{Irreversible spin-transfer and magnetization reversal under spin-injection } 
\author{J.-E. Wegrowe}
\email{jean-eric.wegrowe@polytechnique.fr}
\author{H.-J. Drouhin}
\affiliation{Laboratoire des Solides Irradi\'es, Ecole 
Polytechnique, CNRS-UMR 7642 \& CEA/DSM/DRECAM, 91128 Palaiseau Cedex, France.}

\date{\today} 

\begin{abstract}
 
In the context of spin electronics, the two spin-channel model assumes
that the spin carriers are composed of two distinct populations: the
conduction electrons of spin up, and the conduction electrons of spin
down.  In order to distinguish the paramagnetic and ferromagnetic
contributions in spin injection, we describe the current injection
with four channels : the two spin populations of the conduction bands
($s$ or paramagnetic) and the two spin populations of the more 
correlated electrons ($d$ or ferromagnetic).  The redistribution of 
the conduction electrons at
the interface is described by relaxation mechanisms between the
channels. Providing that the $d$ majority-spin
band is frozen, $s-d$ relaxation essentially concerns the
minority-spin channels.  Accordingly, even in the abscence of
spin-flip scattering (i.e. without standard spin-accumulation or
giant magnetoresistance), the $s-d$ relaxation leads to a $d$
spin accumulation effect.  The coupled diffusion equations for
the two relaxation processes ($s-d$ and spin-flip) are derived.  The 
link with the ferromagnetic order parameter $\vec{M}$ is performed by assuming 
that only the $d$ channel contributes to the Landau-Lifshitz-Gilbert 
equation. The effect of
magnetization reversal induced by spin injection is explained by these
relaxations under the assumption that the spins of the conduction 
electrons act as
environmental degrees of freedom on the magnetization. 

\end{abstract}

\pacs{PACS numbers: 75.40. Gb, 75.70. Pa, 75. 70. Cu \hfill}

\maketitle


Spin-dependent transport in intermetallic ferromagnetic materials is
usually described in the framework of the two spin-channel
approximation.  In this model the spin carriers are composed of two
distinct populations: the conduction electrons of spin up
($\uparrow$), and the conduction electrons of spin down
($\downarrow$), which are defined by their conduction properties
\cite{twochannel}. This model can be applied to ferromagnetic junctions
and to hetero-junctions and it provides simple and operational descriptions
of many effects related to spin dependent transport.  The giant
magnetoresistance ($GMR$) effect
\cite{Gijs,Buttler,Levy0,Valet,LevyDmu} is described by the diffusive
spin-accumulation mechanism in normal / ferromagnetic junctions
\cite{LevyDmu,Johnson1,Wyder,spininjFert,Levy2001} and heterojunctions
\cite{spininjsupra,spininjSC} and also by interface scattering and/or
band mismatch effects.  Beyond the well-known $GMR$, which is due to 
the action of the magnetization on the spin of the conduction electrons,
the opposite effect of the spin polarized current on the
magnetization, the so called "current induced magnetization switching"
or $CIMS$
\cite{Sloncz,Berger,PRBThermo,Tsoi0,EPL,science,Albert,Sun,Kent,Julie}
has also been treated in the framework of the two channel
approximation (see the refrences \cite{jianwei} for the
effects related to noncollinear magnetization).

Tremendous attention has been paid to the microscopic description of
$CIMS$ in terms of the dynamics of the spin of the conduction
electrons in ferromagnets
\cite{Bazaliy,Waintal,Levy2,Stiles,Bauer,Heide,Polianski}.  These approaches add
a deterministic correction ("the spin torque") to the
Landau-Lifshitz-Gilbert ($LLG$) equation due to the current, by
identifying the spins of conduction electrons to the ferromagnetic
order parameter.  The old and unconfortable discussion about
paramagnetic vs.  ferromagnetic character of itinerant spins (already
problematic at equilibrium as discused in the seventeen \cite{Herring}) 
is implicitly evacuated.  A carefull study, however, of the
different kinds of experiments performed about $CIMS$
\cite{SPIE,APL,MSU,Fabian,Tsoi,Derek,Guittienne,Marcel,
PRLStiles,Kiselev,Rippard,Pufall,Urazhdin2,Rque} motivated in parallel
the developpment of a {\it stochastic approach}, where the action of
the spins of the conduction electrons on the magnetization is not
direct, and is described in terms of environmental degrees of freedom. 
The magnetic order parameter and the spins of conduction 
electrons have not the same physical meaning.
The most typical stochastic effect is the two level fluctuation and
activation processes observed only under current injection over a time
window range of a few nanoseconds \cite{Pufall,Guittienne,Fabian,SPIE}
to minutes \cite{Albert}.  Below the nanosecond time range, time
resolved measurements (i.e. one-shot measurements) are not available
but absorbtion peaks are measured in the electric power-spectrum at
frequencies between 3 to 20 GHz \cite{Kiselev,Rippard,Zhu0}.  These
resonances can be though of as a reminiscence of what is observed in
ferromagnetic resonance experiments, or noise measurements without
spin-injection \cite{Russek,Bertram} :at these time scales only
(sub-nanosecond), thermal spin-waves and precession of the
magnetization are expected.  This is the typical time scale of the
dynamics of the magnetization.  At the smaller coarse-grained level,
electronic relaxations take place at times scales ranging from
$10^{-12}$ to $10^{-16}$ second.  These well separate time scales
justify an approach in terms of Markovian processes in which the
relevant slow variable is the ferromagnetic order parameter, and the
fluctuations come from the relaxation of electronic spin degree of
freedom.

The aim of this paper is to propose a phenomenological description of
a possible mechanism of electronic relaxation (presented in Section I)
which accounts for {\it the fondamental difference between paramagnetic
and ferromagnetic spin currents}.  In contrast to the paramagnetic
current associated to the standard spin accumulation process and
$GMR$, which does not interact direcly with the magnetization, the
ferromagnetic, or $d$ channel current (the denomination "$d$
channel" is a reference to the equilibrium description of itinerant $d$ electrons
\cite{Stearns,Ting}) contributes to the ferromagnetic order parameter. 
The relaxation from paramagnetic channels ($s$ channels) to the
ferromagnetic channels ($d$ channels), described as a diffusion process
at the interfaces, leads to spin-transfer from injected current to the
$d$ current, through a new $d$ spin accumulation process
(Sections I and II).  It is furthermore assumed that even this $d$
contribution to the dynamics of the magnetization is not direct, due
to the difference in the typical time scales.  The $d$ current
injection is accounted for in terms of magnetic fluctuations or noise
(Section III) and the after effect is to excite a large spectrum of
magnetic and non-magnetic modes in the ferromagnetic layer.

This approach accounts for both the measured stochastic processes of
the magnetization with a cut-off frequency beyond one GHz, and the
observation of magnetization reversal by spin injection measured in
single uniform magnetic layers (i.e. in the abscence of giant
magnetoresistance) \cite{SPIE,Guittienne,Marcel,PRLStiles}.  The
approach adopted here is based on general arguments from irreversible
thermodynamics, following the general method of De Groot and Mazure
\cite{DeGroot} or Stuekelberg \cite{Stuck}.

Following the pioneering ideas of Mott \cite{Mott}, we assume that two electronic
populations, referred to as $s$ and $d$ are relevant in order to describe the
conduction properties of the ferromagnetic $3d$ metals and alloys. 
This hypothesis was used in order to describe the anisotropic
magnetoresistance in these compounds \cite{Potter}, and to account 
for the contribution to the ferromagnetism of itinerant $d$ electrons 
in equilibrium
\cite{Stearns,Ting} .  In our context,
the $s$ population describes the conduction electrons injected at the
interface of a ferromagnetic nanostructure, and the $d$ population
describes the more localized electrons and associated spins, mainly
responsible for the ferromagnetism of $d$ ferromagnets. 
The simplest extension of the two-channel approximation to our context
is to divide both the $s$ and $d$ populations into up ($\uparrow$) and
down ($\downarrow$) spin populations ($\updownarrow$ is the internal
variable describing the spin degree of freedom as defined in Ref.
\cite{Prigogine}).

Providing that the $d$ majority spin band ($\uparrow$) is fully
occupied, $s-d$ relaxation of $\uparrow$ spin is negligible, and
spin-conserved $s-d$ relaxation only concerns the minority spin
channels $\downarrow$.  If we assume further that $s-d$ relaxation
with spin-flip is negligible with respect to spin conserved $s-d$
relaxation, a new mechanism of spin injection could then be described
as parallel to the usual $ \uparrow - \downarrow $ spin-flip
relaxation used to describe spin accumulation and $GMR$ phenomena
\cite{Valet,spininjFert,Levy2001}.

\section{Thermokinetic equations}

The system is composed by the reservoirs of the injected $s$ electrons
and the ferromagnetic layer composed by the $d$ electrons.  At the
interface, current injection leads to a redistribution of the
different electronic populations that are governed by spin 
polarization and charge
conservation laws.  Let us assume that the current injected is spin
polarized in the down polarization ($\downarrow$).  The conservation
laws should be written by taking into account the reaction mechanisms
between the different populations. At short time scales (electronic 
scattering) the relaxation channels are assumed to be the 
following four \\ \\ \\ 
 (I) $e_{s \downarrow} \to e_{d \downarrow}$ (spin-conserved $s$-$d$ scattering) \\
 (II) $e_{s \downarrow} \to e_{s \uparrow}$ (spin-flip scattering for 
 the $s$ population) \\
 (III) $e_{s \downarrow} \to e_{d \uparrow}$ (spin-flip 
 $s$-$d$ scattering) \\
 (IV) $e_{d \downarrow} \to e_{d \uparrow}$ (spin-flip 
 scattering for the $d$ population) \\
 \\
 
 In a second step (at larger time scales), opposite relaxation
 mechanisms take place (involving ferromagnetic excitations) that are
 described in Sec. III. Process (I) is assumed to be the main
 mechanism responsible for irreversible spin transfer.  Process (II)
 leads to the well-known spin-accumulation effect and was described in
 detail with the same formalism elsewhere \cite{PRBThermo}.  {\it
 According to the fact that the majority-spin $d$ band is full} and
 lies at a sizable energy below the Fermi level, the current $J_{d
 \uparrow}$ is negligible and the channel $d \uparrow $ is frozen. 
 Processes (III) and (IV) are hence negligible \cite{Drouhin}. 
 Consequently, we are dealing with a three channel approximation.
 
  The total current $J_{t}$ is composed by the three currents for each 
 channel $\{ s\uparrow, s \downarrow, d \downarrow \} $:
 $J_{t} = J_{s \uparrow}+J_{s \downarrow}+J_{d 
 \downarrow}$.
 In order to write the conservation laws, the 
 relaxation rate $\dot \Psi_{sd} $, is introduced to
account for $s-d$ spin-conserved scattering, and the relaxation rate
$\dot \Psi_{s} $, is introduced in order to
account for spin-flip scattering. 
 Assuming that the current is flowing along the z axis, the 
 conservation laws for a steady state regime are :

 \begin{equation}
   \begin{array} {lllll}
  \frac{d J_{t}}{dt} =  - \frac{\partial J_{t}}{\partial z} = 0 \\
\frac{d J_{s \uparrow}}{dt}\, = \,-\frac{\partial J_{s \uparrow}}{\partial 
z} - \, \dot{\Psi}_{s} = 0 \\ 
\frac{dJ_{s \downarrow}}{dt}\, = \,-\frac{\partial J_{s \downarrow}}{\partial z} - \, \dot{\Psi}_{sd} 
 + \, \dot{\Psi}_{s} = 0\\
\frac{dJ_{d \downarrow}}{dt}\,  =  -\frac{\partial J_{d \downarrow}}{\partial z} + 
\dot{\Psi}_{sd} = 0
\end{array} 
\label{con}
\end{equation}

The system is described by the number of
electrons present in each channel at a given time, that defines the
four currents, plus the entropy of the system.  The conjugated
(intensive) variables are the chemical potentials $\{ \mu_{s
\uparrow}, \mu_{s \downarrow}, \mu_{d \uparrow}, \mu_{d \downarrow }\}$. 
As described in
Appendix A, the
application of the first and second laws of thermodynamics allows us 
to deduce the
Onsager relations of the system :

\begin{equation}
\begin{array} {lllll}
J_{s \downarrow} = -\frac{\sigma_{s \downarrow}}{e} \frac{\partial 
\mu_{s \downarrow}}{\partial z}\\ 
J_{s \uparrow} = -\frac{\sigma_{s \uparrow}}{e} \frac{\partial \mu_{s 
\uparrow}}{\partial z}\\ 
J_{d \downarrow} = -\frac{\sigma_{d \downarrow}}{e} \frac{\partial 
\mu_{d \downarrow}}{\partial z}\\ 
\dot{\Psi}_{sd} = L_{sd} \left ( \mu_{s 
\downarrow}-\mu_{d \downarrow} \right ) \\ 
\dot{\Psi}_{s} = L_{s} \left ( \mu_{s \uparrow}-\mu_{s  \downarrow} \right )
\end{array}
\label{Onsager0}
\end{equation}

where the conductivity of each channel $ \{ \sigma_{s \uparrow}, 
\sigma_{s \downarrow}, \sigma_{d \uparrow}, \sigma_{d \downarrow} \}$ 
has been introduced. 
The first four equations are nothing but Ohm's law applied to each
channel, and the two last equations introduce new Onsager transport
coefficients (see Appendix A), $L_{sd \downarrow}$ and $L_{s}$, that 
respectively describe the
$s-d$ relaxation (I) for minority spins under the action of the chemical potential
difference $\Delta \mu_{\downarrow} = \mu_{s \downarrow}-\mu_{d
\downarrow}$ and the spin-flip relaxation (II) under spin pumping 
$\Delta \mu_{s} = \mu_{s \uparrow}-\mu_{s \downarrow}$. 

The quantities of physical interest are 
the {\it paramagnetic} current $J_{0s}= J_{s \uparrow} + J_{s 
\downarrow}$, the minority-spin current  $ J_{0 \downarrow} = J_{s
\downarrow} + J_{d \downarrow}$, and the two {\it polarized currents} $ \delta J_{\downarrow} = J_{s 
\downarrow} - J_{d \downarrow}$ and $\delta J_{s}= J_{s \uparrow} - 
J_{s \downarrow}$. We introduce the $\sigma_{s}$ and $\sigma_{\uparrow}$
conductivities $ \{ \sigma_{s} = \sigma_{s \uparrow} + \sigma_{s 
\downarrow} $ and 
$\sigma_{\downarrow} = \sigma_{s \downarrow} + \sigma_{d 
\downarrow} \}$. The conductivity imbalance $ \beta_{ \downarrow}¥ $ and $ \beta_{s}¥ $ between 
respectively the $s \downarrow$ and $d \downarrow$ channels and the $s \uparrow$ and $s \downarrow$ 
channels are: 

\begin{equation}
  \begin{array} {ll}
\beta_{\downarrow}¥ = \frac{ \sigma_{s \downarrow}-\sigma_{d
\downarrow}}{\sigma_{\downarrow} }\\
\beta_{s}¥ = \frac{ \sigma_{s \uparrow}-\sigma_{s
\downarrow}}{\sigma_{s}}
\end{array} 
\label{beta}
\end{equation}

Eqs.~(\ref{con}) becomes :
\begin{equation}
 \begin{array} {lllll}
\frac{\partial J_{t}}{\partial z} =  \frac{\partial J_{
d \downarrow}}{\partial z} + \frac{\partial J_{s}}{\partial z} = \, 0 \\
\frac{\partial J_{0 \downarrow}}{\partial z} =  \, \dot{\psi}_{s} \\
\frac{\partial \delta J_{\downarrow}¥}{\partial z}\, = \, -2 
\dot{\psi}_{sd} - \dot{\psi}_{s}¥ \\
\frac{\partial J_{0s}}{\partial z} =  \, - \dot{\psi}_{sd} \\
\frac{\partial \delta J_{s}¥}{\partial z}\, = \,
\dot{\psi}_{sd} - 2 \dot{\psi}_{s}
\label{cont}
\end{array} 
\end{equation}

and, defining the quasi-chemical potentials \cite{Parrott} $\mu_{s} = 
\mu_{s \uparrow}+\mu_{s \downarrow}$ and  $\mu_{\downarrow} = 
\mu_{s \downarrow}+\mu_{d \downarrow}$,
Eqs.~(\ref{Onsager0}) becomes :

\begin{equation}
   \begin{array} {lllll}
J_{0 \downarrow} = -\frac{\sigma_{\downarrow}}{2e} \left ( \frac{\partial \mu_{\downarrow} }{\partial 
z} + \beta_{\downarrow}¥ \frac{\partial \Delta \mu_{\downarrow}¥}{\partial 
z}  \right ) \\ 
\delta J_{\downarrow}¥ = -\frac{\sigma_{\downarrow}}{2e} \left (   \beta_{\downarrow}¥ 
\frac{\partial \mu_{\downarrow}}{\partial z} + \frac{\partial \Delta 
\mu_{\downarrow}¥}{\partial
z} \right ) \\
J_{0s} = -\frac{\sigma_{s}}{2e} \left ( \frac{\partial \mu_{s} }{\partial 
z} + \beta_{s}¥ \frac{\partial \Delta \mu_{s}¥}{\partial 
z}  \right ) \\ 
\delta J_{s}¥ = -\frac{\sigma_{s}}{2e} \left (   \beta_{s}¥  \frac{\partial
\mu_{s}}{\partial z} + \frac{\partial \Delta \mu_{s}¥}{\partial
z} \right ) \\
\dot{\Psi}_{sd} = L_{sd} \Delta \mu_{\downarrow} \\
\dot{\Psi}_{s} = L_{s} \Delta \mu_{s}
\end{array}  
\label{Onsager1}
\end{equation}

 The equations of
conservation [Eqs.~(\ref{cont})] and the above Onsager equations lead to
the two coupled diffusion equations :

\begin{equation}
 \begin{array} {ll}
\frac{\partial^2 \Delta \mu_{\downarrow}¥}{\partial z^2}\,=\, 
\frac{1}{l_{sd}^2} \, \Delta 
\mu_{\downarrow}- \frac{1}{\lambda_{s}¥^{2}} \Delta \mu_{s} \\
\frac{\partial^2 \Delta \mu_{s}¥}{\partial z^2}\,=\, 
\frac{1}{\lambda_{sd}^2} \, \Delta 
\mu_{\downarrow}- \frac{1}{l_{sf}^2} \, \Delta 
\mu_{s}
\end{array} 
\label{diff}
\end{equation}

where 

\begin{equation}
 \begin{array} {llll}
l_{sd}\,\equiv \, \sqrt{\frac{ \sigma_{\downarrow} 
\left ( 1 - \beta_{\downarrow}¥^{2}\right )}{4¥eL_{sd}}} \\
\lambda_{s} \equiv \, \sqrt{\frac{ \sigma_{\downarrow} 
\left ( 1 + \beta_{\downarrow} \right )}{2eL_{s}}} \\
l_{sf}\,\equiv \, \sqrt{\frac{ \sigma_{s} \left ( 1 - \beta_{s}¥^{2}\right 
)}{4eL_{s}}} \\
\lambda_{sd}\,\equiv \, \sqrt{\frac{ \sigma_{s} 
\left ( 1 - \beta_{s}
\right )}{2eL_{sd}}}
\end{array}   
\label{ldiff}
\end{equation}

A solution of Eqs. (\ref{diff}) is 

\begin{equation}
\begin{array} {ll}
\Delta \mu_{\downarrow}\,=\, \Delta \mu_{1} + \Delta \mu_{2} \\
\Delta \mu_{s}\,=\, \lambda_{s}^{2} \left ( 
\left ( \frac{1}{l_{sd}^{2}} - \frac{1}{\Lambda_{+}^{2}} \right) \, \Delta \mu_{1} 
+  \left ( \frac{1}{l_{sd}^{2}} - \frac{1}{\Lambda_{-}^{2}} \right ) \Delta \mu_{2}  \right )
\end{array}
\label{soldiff}
\end{equation}

with 

\begin{equation}
\begin{array} {ll}
\Delta \mu_{1} \,=\, a_{1}¥ e^{\frac{z}{\Lambda_{+}}} + a_{2}¥ e^{- \frac{z}{\Lambda_{+}}} \\
\Delta \mu_{2} \,=\, b_{1}¥ e^{ \frac{z}{\Lambda_{-}}} + b_{2}¥ e^{- \frac{z}{\Lambda_{-}}}
\end{array}
\label{delta}
\end{equation}

where
$$\Lambda^{-2}_{\pm} = \frac{1}{2}(l_{sd}^{-2}+l_{sf}^{-2}) 
\left ( 1 \pm \sqrt{1-4 \frac{ l_{sd}^{-2}l_{sf}^{-2}
-\lambda^{-2}_{s}\lambda_{sd}^{-2}}{ \left ( l_{sd}^{-2}+l_{sf}^{-2} 
\right )^{2} }} \right ) $$

The constants $a_{1}¥$, $a_{2}¥$, $b_{1}¥$, $b_{2}¥$ are defined by the boundary conditions.
 It can then be seen that the usual spin accumulation corresponding to $\Delta
 \mu_{s}$ also depends on the
 spin-conserved $s-d$ electronic diffusion which is known to be 
 efficient \cite{Drouhin} and, conversely, that
 spin-conserved diffusion is able to lead to a spin accumulation,
 or {\it $d$ spin-accumulation} effects.  Accordingly, we 
 expect to measure some typical effects related to spin-accumulation
 in single magnetic layers, or if $\beta_{s} = 0$ : this point will be
 illustrated in the new expression of the magnetoresistance 
 (Eq. (\ref{GMR}) below), and in Section III through the effect of
 $CIMS$.  $s-d$ relaxation adds a new
 contribution to the resistance, which plays the role of an interface
 resistance arising from the diffusive treatment of the band 
 mismatch \cite{Gijs,Buttler,Levy0,Valet,LevyDmu}.  
 
 The resistance produced by the spin-flip contribution (usually defined
 as the giant magnetoresistance $R_{GMR}$), plus the contribution of 
 $s-d$ relaxation, are defined
 by 

 \begin{equation}
	 R_{GMR} = \frac{1}{J_{t}} \int_{Ð\infty}^{+ \infty} \left (
	 \frac{-1}{e} \frac{\partial \mu_{t} }{\partial z} - E_{t}^{\infty} \right ) dz
	 \end{equation}

 where $E_{t} = -\frac{\partial \mu_{t}}{e \partial z}$ is the total
 electric field and $E_{t}^{\infty}$ is the electric field far away
 from the interfaces.  Providing that the total current is $J_{t} =
 J_{s \uparrow}+J_{s \downarrow}+J_{d \downarrow}$, or
 
  \begin{equation}
 J \, = \, -\frac{\sigma_{t}}{e} \frac{\partial}{\partial z} \left ( \frac{\sigma_{d 
 \downarrow}¥}{\sigma_{t}} \,  \mu_{d \downarrow}¥+  \frac{\sigma_{s 
 \downarrow}¥}{\sigma_{t}} \, \mu_{s \downarrow} + \frac{\sigma_{s 
 \uparrow}¥}{\sigma_{t}} \,  \mu_{s \uparrow} \right )
 \label{current}
 \end{equation}
 
 The total electric field can also be written (from Eqs. (\ref{Onsager0})) 
 as 
 
 \begin{equation}
 E_{t} \, = \, -\frac{\partial \mu_{t}}{e \partial z} = \frac{J_{t}}{\sigma_{t}}=
 -\frac{1}{e} \left ( \frac{\partial \mu_{d \downarrow}}{\partial z} +
 \frac{\sigma_{s}}{\sigma_{t}} \frac{\Delta \mu_{\downarrow}}{\partial
 z} + \frac{\sigma_{s \uparrow}}{\sigma_{t}} \frac{\Delta \mu_{s}}{\partial
 z} \right )
 \end{equation}
 
 where $\sigma_{t}=\sigma_{s \uparrow} + \sigma_{s \downarrow} +
 \sigma_{d \downarrow}$ , and $E_{t}^{\infty} = lim_{z \rightarrow + 
 \infty} \, ¥-\frac{1}{e}
 \frac{\mu_{d \downarrow}}{\partial z}$.  The resistance is given by :

 \begin{equation}
	 R_{GMR} = - \frac{1}{e J_{t}} \int_{- \infty}^{+ \infty}¥\left (
	 \frac{\sigma_{s}}{\sigma_{t}} \frac{\partial \Delta
	 \mu_{\downarrow}}{\partial z} + \frac{\sigma_{s
	 \uparrow}¥}{\sigma_{t}}\frac{\partial \Delta
	 \mu_{s}}{\partial z} \right )dz
	 \label{GMR}
\end{equation}
 
This three-channel model
brings to light the interplay between band mismatch effects and spin
accumulation, in a diffusive approach.  It is interesting to note that
the local neutrality charge condition which is often used (see for
instance Eq. (4) in \cite{Rashba}) was not included.  On the contrary, we have
imposed the conservation of the current at any point of the conductor. 
Indeed, electron transfer from a channel to another where the electron
mobility is different, induces a local variation of the total current.

\section{Normal - ferromagnetic interface}

The resolution of Eqs.~(\ref{diff}) leads to a variety of possible 
behaviours, from single interface effects to superlattice effects 
(see the paper by Valet -Fert \cite{Valet} for the discussion in the 
framework of the two-channel approximation). Our main goal however is to 
understand the contributions of standard spin-accumulation and 
$d$ spin-accumulation in uniform 
magnetic layers where no $GMR$ are present due to the symmetry between 
both interfaces.  We first focus 
our attention on a single 
interface separating two semi-infinite layers. If the ferromagnetic 
layer is at the right hand side, the solutions follow, from Eqs.~(\ref{soldiff})

\begin{equation}
\begin{array} {ll}
\Delta \mu_{1} ( z) \,=\, b e^{ - \frac{z}{\Lambda_{+}}} \\
\Delta \mu_{2} ( z) \,=\, d e^{ - \frac{z}{\Lambda_{-}}} 
\end{array}
\label{delta2}
\end{equation}

In our context, we define a "normal" metal as a compound with fully 
occupied $d$ bands (the $d
\uparrow$ and $d \downarrow$ channels are frozen).  In the normal
metal (on the left hand side of the junction) we have $\Delta
\mu_{s}^{N} = a e^{-z/l_{sf}}$.

At the junction, the continuity of the currents and the continuity of $\Delta 
\mu_{s}$, without interface resistance, becomes :

 \begin{equation}
\begin{array} {lll}
\left ( \frac{\partial \mu_{d \downarrow}¥}{\partial z} \right )_{0^{+}} = 0 \\
J_{s \updownarrow}^{N}(0^{-})= J_{s \updownarrow }^{F}(0^{+}) \\
\Delta \mu_{s}^{N}(0^{-}) = \Delta \mu_{s}^{F}(0^{+})
\end{array}  
\label{condlimite}
\end{equation} 

Where the superscripts $N$ and $F$ resp. stand for normal and ferromagnetic
metals. The system originated from Eqs.~ (\ref{condlimite}) is solved
in Appendix B. The three limiting cases of special interest are
presented below.

Let use first consider the standard assumption where $GMR$ and
spin-accumulation are calculated.  This case corresponds to both
limits $l_{sd} \ll l_{sf}$ (i.  e. instantaneous $s - d$ relaxation at
the interface), or $l_{sf} \ll l_{sd}$ (no $sd$ relaxation in the time
scale of spin-flip relaxation).  The simple case of two identical
adjacent layers with opposed spin polarizations is generally
considered \cite{Valet}.  In our context, this situation would
correspond to a "quasi-ferromagnetic" metal in which both $d \uparrow$
and $d \downarrow$ are frozen (i.  e. there is no $d$ ferromagnetism)
but with $\beta_{s}^{N}=-\beta_{s}¥^{N}$, i. e. there is nevertheless a
spin polarization of the current.  The well-known solution is
straightforwardly recovered.  From the equations presented in Appendix
B:

\begin{equation}
\Delta \mu_{s}¥ \approx  \frac{2 e \beta_{s} l_{sf}}{\sigma_{s} 
(1-\beta_{s}^{2}¥)} J_{t}
e^{-z/l_{sf}}
\label{Dmus2}
\end{equation}

In the case of a junction with a "normal" metal $\beta_{s}^{N}=0$ and
a ferromagnetic metal $\beta_{s} \ne 0$, the well-known
result is also recovered in both limits (with 
$\sigma_{s}^{N}=\sigma_{s}^{F} =
\sigma_{t}$ and $l_{sf}^{N} \approx l_{sf}^{F}$) :

\begin{equation}
\Delta \mu_{s}¥ \approx \frac{2 e \beta_{s} l_{sf}}{\sigma_{s} 
\left (1-\frac{\beta_{s}^{2}}{2} \right )}  J_{t}
e^{-z/l_{sf}}
\label{Dmus2bis}
\end{equation}

in these two limits there is hence no modification due to the 
existence of the
"$d$ spin-accumulation" of the standard spin accumulation $\Delta 
\mu_{s}$. The $d$ spin-accumulation however is not zero. In the 
case of a normal metal / ferromagnetic junction with $\sigma^{F} = 
\sigma^{N}$ , the 
$d$ spin-accumulation is :

\begin{equation}
\Delta \mu_{\downarrow}¥ \approx \frac{e l_{sd}}{\sigma_{s}} 
\frac{ \left ( 1 + \frac{\beta_{s}}{2} \right ) } { \left ( 1 - 
\frac{\beta_{s}^{2}}{2} \right )} J_{t} 
e^{-z/l_{sd}}
\label{Dmudown}
\end{equation}

  The corresponding contribution to the magnetoresistance is
  proportionnal to $l_{sd}^{2}$ and $\frac{1}{1-\frac{\beta_{s}^{2}}{2}}$.  This
  contribution is not zero even if the spin polarization of the
  current is vanishing ($\beta_{s}=0$) in the ferromagnet.  
 Since $l_{sd}$ is however expected to be very small, the
  $d$ spin-accumulation contribution should be important
  only for $\beta_{s}$ close to $\sqrt{2}$ (this is a consequence of 
  the assumption $\sigma^{F} l_{sf}^{N} = \sigma^{N} l_{sf}^{F}$).
  
  In intermediate cases, when $l_{sf}$ is of the same order of 
  magnitude as $l_{sd}$, the spin-accumulation $\Delta \mu_{s} $ is 
  non-vanishing even if $ \beta_{s} $ tends to zero. 

\section{Generalized Landau-Lifshitz-Gilbert equation}

The question related to magnetization reversal under spin-injection is
how the kinetics of the spins of the conduction electrons described in
the previous sections is related to the ferromagnetic order parameter.  In order to
investigate this problem, we shall consider the different quantities
that define our system (spins of conduction electron, electric
charges, magnetization \ldots), and the corresponding relevant scales,
or {\it coarse-graining } \cite{Foster,Balian,Haenggi}.

In our experimental context, the magnetization is a collective
variable whose dynamics are much slower than all other paramagnetic
spin relaxation mechanisms because the magnetization is conserved over
the distance of the magnetic layer (or exchange length for spin
waves), while the conservation of the electric charges is relevant
over a local equilibrium of the order of the nanometer.  It is
then possible to identify three different time-scales. 
The first is the electronic relaxation ($ 10^{-15}$ to $ 10^{-12}$ for 
paramagnetic transvers spin effects \cite{Levy2}),
the second is the typical ("quasi-ballistic") dynamics of the
magnetization ($10^{-11}$ to $10^{-9}$), and finally, the time scale of the
activation processes measured over decades from
10 nanoseconds to hours.  The electronic degrees of freedom, and especially the
spins of the four electron populations defined in the first section
can then be treated in terms of the action of an environment in a
stochastic approach (e. g. defining a projection operator over the
relevant variable $ \vec{M}_{0}$ \cite{Foster,Balian,Haenggi}).  The
effects of the spins of the conduction electrons are then reduced to
the noise and the damping coefficient.  Without current injection
\cite{PRBThermo}, the electronic degrees of freedom are contained in
the Gilbert damping term \cite{Kamberski,Ho}, and the following
Gilbert equation for the magnetization $ \vec{M}_{0}$ of the
ferromagnetic layer is obtained \cite{Coffey}:

\begin{equation}
\frac{d \vec{M}_{0}}{dt}\,= \, \Gamma 
\, \vec{M}_{0}\, \times \, \left \{ 
-\frac{\partial V}{\partial \vec{M}_{0}}\,-\, \eta \, \frac{d 
\vec{M_{0}}}{dt} \right \}
\label{dynamicsM}
\end{equation}
 
where $\Gamma$ is the gyromagnetic ratio and $V$ is the magnetic Gibbs
potential \cite{Brown}, where $\vec{H}^{ext}$ being the external magnetic
field.  The $\eta$ coefficient is the Gilbert damping factor. 
Eq.~(\ref{dynamicsM}) can also be put into the following
Landau-Lifshitz form.  In the case of uniform magnetization we have
$\vec{M}_{0}\,=\, M_{s}\vec{u}_{0}$, where $M_{s}$ is the saturation
magnetization and $\vec{u}_{0}$ the unit vector, so that it
becomes:

\begin{equation}
\frac{d \vec{u}_{0}}{dt}\,= \,-\,g'\left(\vec{u}_{0} \times 
\vec{\nabla} V \right) -h' \vec{u}_{0} \times \left(\vec{u}_{0} \times 
\vec{\nabla} V\right)
\label{LL}
\end{equation}

where $\vec{\nabla} $ is the gradient operator on the surface of a unit sphere.
The phenomenological parameters h' and g' 
are linked to the gyromagnetic ratio $ \Gamma $ 
 and the Gilbert damping coefficient $ \eta $ by the relations 

\begin{eqnarray}
\left\{ \begin{array}{ccc}
 h'\,&=&\,\frac{\Gamma \alpha}{(1+\alpha^2)M_{s}^{2}¥} \nonumber \\
g'\,&=&\,\frac{\Gamma}{(1+\alpha^2)M_{s}} \nonumber \\
\alpha\,&=&\, \eta \Gamma M_{s}
\end{array} \right . 
\end{eqnarray}

If we take into account the relaxation processes (I) to (IV), the
conservation of the ferromagnetic spins normal/ferromagnetic
interface is:

\begin{equation}
\dot{\vec{u}}\,= \,-\,g'\left(\vec{u} \times 
\vec{\nabla} V \right) -h' \vec{u} \times \left(\vec{u} \times 
\vec{\nabla} V\right) \, - \, f_{\|}¥(t) \vec{u} 
\label{GLL}
\end{equation}

where 

\begin{equation}
	f_{\|}(t) = \frac{g_{d} \mu_{B}}{g \mu_{B} L} \int \left ( \delta J_{
\downarrow}^{int}(z) - \delta J_{
\downarrow}^{out}(z)  \right )dz 
\end{equation}

$L$ is the length of the magnetic layer, and $g_{d} \mu_{B}$ is the
magnetization per atom (number of Bohr magnetons) for the $d$
population; $g \mu_{B}$ is the magnetization per atom of the
ferromagnet, and the superscripts $int$ and $out$ describe
respectively the interface for the incoming current, and the
interface corresponding to the current flowing out of the layer.  The
$\delta J_{\downarrow}$ current is defined in Eq.(\ref{Onsager1}) :

\begin{equation}
\delta J_{\downarrow}¥ = -\frac{\sigma_{\downarrow}}{2e} \left (   \beta_{\downarrow}¥ 
\frac{\partial \mu_{\downarrow}}{\partial z} + \frac{\partial \Delta 
\mu_{\downarrow}¥}{\partial
z} \right )
\end{equation}

with the solutions given in Section II and Appendix B. Note that
$\vec{u} $ is no longer a unit vector due to the last term in the
right hand side of the equation Eq.  \ref{GLL}.  Instead, $\vec{u} $
includes the fluctuations of $\vec{u}_{0}¥$. After projection over
$\vec u$, Eq.  (\ref{GLL}) gives the variation of the modulus of the
magnetization due to current injection (\cite{Argu}):

\begin{equation}
\frac{d \| u \|^{2} (t)}{dt}\,=  - \frac{g_{d}}{gL} \| u \|^{2} 
\int_{-\infty }^{+ \infty} \left ( \delta J_{
\downarrow}^{int}(z) - \delta J_{
\downarrow}^{out}(z)  \right ) \, dz
\label{module}
\end{equation}

and 

\begin{equation}
\| u \|^{2} (t)\,= \| u _{0}¥\|^{2} e^{- \left \{ \frac{g_{d}}{gL}
\int_{0}^{L} \left ( \delta J_{
\downarrow}^{int}(z) - \delta J_{
\downarrow}^{out}(z)  \right ) \, dz  \right  \} t }
\label{sol}
\end{equation}

 The integral in the exponential is rather similar to that
 present in the calculation of the $GMR$, except that the current is 
 the ferromagnic spin current.  For symmetry reasons
 equal and opposite electronic relaxations are expected to occur at both interfaces of
 the ferromagnetic layer : the effect of spin-injection should be
 compensated : $\int \left ( \delta J_{ \downarrow}^{int}(z) - \int
 \delta J_{ \downarrow}^{out}(z) \right ) dz = 0$.  We may still
 conclude that, if $d \downarrow$ spins do not interact with the
 magnetization and if the two interfaces of the ferromagnet are
 totally symmetric, the spin-transfer would vanish.

This is indeed what happends in the case of $s$ population and
$GMR$, where the energy due to the spin-flip is tranferred into the lattice
and dissipated into the heat bath. This is simply due to the thermalization 
associated to paramagnetism. To 
that respect, the idea of conservation of momenta 
which leads to spin-transfer theories is not 
valide for paramagnetic spins, because the momentum dissipates into the 
heat bath and is not transfered
into the ferromagnetic order parameter. 
In Eq. (\ref{GLL}), the transvers stochastic force
responsible for the fluctuations $f_{\perp}(t)$ has been put to zero
because of the averaging over the equilibrium distribution ($ \langle 
f_{\perp}(t) \rangle_{0}¥ =0$): the transvers
stochastic force is suppose to be the same with or without current 
injection and $ \langle 
f_{\perp}(0) f_{\perp}(t) \rangle_{0} = c \eta kT \delta(t)$ where T is the 
temperature of the lattice and $c= \Gamma M_{s}$ \cite{Coffey}. 

In contrast, the energy dissipated or gained by $s-d $ scattering is
tranfered in the form of a contribution to the ferromagnetic order
parameter within a typical time scales of the dynamics of the
magnetization, and is not dissipated into the heat bath.  Thus, the $d
\downarrow $ spins do interact with the magnetization within a short
distance.  The process described in Eq.  (\ref{sol}) can only be
observed locally, and in the sub-nanosecond time range.  Beyond, the
behavour of the magnetic layer will reflect the large spectrum of
relaxation channels of the magnetization, from spin-waves, solitons,
precession, soft modes etc.  The detailed description of these
contributions is beyond the scope of the present paper, but the
resulting behaviour of the magnetization can be described
phenomenologically in tems of random fluctuations of the magnetization
and effective temperature.  Instead of Eq.  (\ref{sol}), we have
$\langle \| u \|^{2} \rangle= \lim_{t \rightarrow \infty }\| u
\|^{2}(t) \approx 0 $.  At long enough time scales or thick enough
layers, the last term in the right hand side of Eq. (\ref{GLL}) is
averaged out : $$ \langle f_{\|}(t) \rangle \approx 0 $$.

The stationary
transfer of spins, described here as environmental degrees of freedom,
emerges in terms of transfer of energy (and entropy) through the
non-vanishing fluctuations of the modulus of the magnetization.  {\it
The fluctuation-dissipation relation, defines the current dependent
effective temperature } $T_{eff}(J_{t})$ with the correlations:

\begin{equation}
\langle  
f(0)_{\|} f(t)_{\|} \rangle \equiv \tilde{c} \eta \, k T_{eff}(J_{t}) \, \delta(t)
\end{equation}

where k is the Boltzmann constant and $\tilde{c}$ is an appropriate
constant.  The energy $k T_{eff}$ is stored in the layer in form of
magnetic fluctuations, and is not direcly related to the temperature 
of the lattice.  Accordingly, it is possible to transfer some few eV
without damage in a nanoscopic sample (except if the fluctuations are
generating spin-waves only, because spin-waves relax very rapidly into
the lattice) \cite{SPIE}.  It is also expected that the efficiency of
the transfer is maximum if the layer thickness is large.  This would explain the
behaviour as the function of the temperature observed in ultrathin
trilayer nanopillars \cite{Tsoi,Albert}; the amplitude of $CIMS$
decreases while decreasing the temperature, and in long Ni nanowires
\cite{Marcel} where the amplitude of $CIMS$ is constant or increases
while decreasing the temperature.

The effective temperature is proportional to $\delta J_{\downarrow
}^{2}$, and depends, through $\Delta \mu_{s}$ on the spin accumulation
properties, and through $\Delta \mu_{\downarrow}$ to the $d$
spin-accumulation.  The fluctuations of the current $\delta
J_{\downarrow }(t)$ are hence able to account for the effective
temperature measured in some experiments of magnetization reversal
\cite{Guittienne,MSU,Fabian,SPIE} in both single layer and trilayer
systems.  The effect described in terms of effective temperature is
not directly sensitive to the sign of the current, but depends
indirectly on the current direction through the sign of $\Delta
\mu_{s}$ and $\Delta \mu_{\downarrow}$, because it inverses the
asymetry at the interfaces \cite{Valet} in multilayered $GMR$ devices. 
In contrast, the spin accumulation is symetric in a single magnetic
layer so that the asymmetry of the interfaces is not necessarily
modified by inverting the current.  It is then possible to account for
both the dependence to the current direction in trilayer structures
\cite{Albert,Julie,Sun,Kent,MSU,Fabian}, and the absence of dependence
in $AMR$ single layer measurements \cite{Derek,Marcel}.

On the other hand, the resonance measured at the GHz frequency range
\cite{Tsoi,Kiselev,Rippard} may be described by the magnetic
excitations (precession and spin-waves, etc) produced during the relaxation
of the modulus of the longitudinal magnetization $\vec{u} $ into the
ferromagnetic layer, while direct measurements of electronic
resonances should be expected at the 10 to 100 GHz frequency range.

It is not possible to perform a direct estimation of spin transfer at
the present stage of the description because the amplitude of the
fluctuations strongly depends on specific magnetic properties of the
sample and interfaces.  A rough estimate can nevertheless be
performed by observing that the typical current density injected at
about $1.6 \, 10^{7} A/cm^{2}$, which corresponds to $10^{16}$
electrons per second injected at the interface.  The current $J_{0
\downarrow}$ is a fraction of that (lets say above 1 \%
\cite{Stearns}).  The question is to know what is the typical length
l, or the time $\tau$ (l=v $\tau$ ) over which the magnetization is
maintained out of equilibrium in the $d \downarrow $ channel.  The
minimum relaxation time should be about $10^{-12}$ seconds, the maximum
should be the typical magnetization dynamics, around $10^{-9}$
seconds.  This means that the system is pertubated by a magnetization
variation of $10^{2}$ to $10^{5} \mu_{B}$.  Note that $2 .10^{4} \mu_{B}$
corresponds to an energy transferred by the current of more than one
eV in a local ferromagnetic field of 1 Tesla (in agreement with
experimental results \cite{SPIE}).  It is then possible to account for
a transfer of magnetic momentum, with an energy largely beyond the
Curie energy in the local field, and which would produce the
magnetization reversal or magnetic excitations.

\section{conclusion}
A new electronic relaxation mechanism has been proposed
under spin-injection at a normal/ferromagnetic interface.  The
description of the relaxation is based on a three-channel model that
leads to a redistributions of electrons between paramagnetic and
ferromagnetic spin currents at the interfaces.  As a consequence, a
new spin accumulation process of the $d$ electrons occurs.  The coupled
diffusion equations are derived and solved.  The contribution of the
$d$ spin accumulation to the standard spin accumulation in the GMR is 
calculated.  The $d$ spin accumulation adds a new
 contribution to the resistance, which plays the role of the interface
 resistance arrising from the diffusive treatment of the band 
 mismatch. 

In contrast to the paramagnetic current (here associated to the
standard spin accumulation process) which does not interact direcly
with the magnetization, the ferromagnetic, or $d$-channel current
contributes to the ferromagnetic order parameter.  It is furthermore
assumed that even this $d$ contribution to the ferromagnetic order
parameter is not direct, due to the difference in the typical time
scales.  The $d$ current injection is accounted for in terms of
magnetic fluctuations or noise the consequence of which  is to excite a large
spectrum of magnetic and non-magnetic excitations in the ferromagnetic
layer.  This stochastic approach allows an effective temperature (or
equivalently an effective potential barrier) to be defined in
agreement with the experimental observations.  The fluctuations
depends, through the interband current $\delta J_{\downarrow }$, to
both the usual spin accumulation $\Delta \mu_{s}=\mu_{s \downarrow} -
\mu_{s \uparrow}¥$ and the d spin accumulation $\Delta \mu_{\downarrow
}=\mu_{s \downarrow}-\mu_{d \downarrow}¥$.  This mechanism allows the
effect of current induced magnetization switching, including 
current-induce activation, to be described not only
in multilayered structures exhibiting $GMR$, but also in uniformly
magnetized nanostructures measured with $AMR$.

\section{Acknowledgement}
HJD thanks the D\'el\'egation G\'enerale pour l'Armement for support.

\section{appendix A}

  The aim of this Appendix is to derive the Onsager matrix
  (\ref{Onsager0}) on the basis of the first and second laws of
  thermodynamics.  In a typical one dimensional junction the layer is
  decomposed into $\Omega$ parts, defining the sub-system $ \Sigma^k
  $, which is in contact to the ``reservoirs'' $ \Sigma^{k-1} $ and $
  \Sigma^{k+1} $.  The sub-systems $ \Sigma^k $, is then an open
  system which exchanges heat and chemical species with its left and
  right vicinity layers.  Furthermore, the populations ($N^k_{s
  \uparrow}$, $N^k_{s \downarrow}$, $N^k_{d \uparrow}$) and spin down
  ($N^k_{d \downarrow}$) are not conserved due to transitions from one
  channel to the other.

In this picture, the states of the sub-system $ \Sigma^k $
 are described by the variables

\begin{equation}
(S^k, N^k_{s \uparrow},N^k_{s \downarrow}, N^k_{d \uparrow},N^k_{d \downarrow})
\label{var}
\end{equation}

where $S^k$ is the entropy. The internal variables $\Psi_{s}¥$, 
$\Psi_{d}$ and $\Psi_{sd}$ must however be introduced in order to take 
into account the relaxation processes occuring respectively between 
the two s-spin channels, the two d-spin channels, and the s-d 
relaxation .

Let us define the heat and chemical power by $P_{\phi}$ (the mechanical 
power is zero as long as the action of the magnetic field 
on the charge carriers is neglected).  The {\it first law of the 
thermodynamics} applied to the layer $\Sigma^k$ gives

\begin{equation}
\frac{dE^k}{dt}\,=\,P_{\phi}^{k-1 \to k}\,-\,P_{\phi}^{k \to 
k+1}
\end{equation}

Introducing the canonical definitions $T^k=\frac{\partial 
E^k}{\partial S^k} $ and  $ \mu_{s\pm}^k=\frac{\partial 
E^k}{\partial N_{s\pm}^k}, \, \mu_{d\pm}^k=\frac{\partial 
E^k}{\partial N_{d\pm}^k}\,$ the energy variation is:

\begin{equation}
\frac{dE^k}{dt}\,=\,T^k \frac{dS^k}{dt}\,+\, \mu_{s \uparrow}^k \frac{dN_{s\uparrow}^k}{dt}\,
+\, \mu_{s \downarrow}^k \frac{dN_{s \downarrow}^k}{dt} \, + \, \mu_{d \uparrow}^k \frac{dN_{d \uparrow}^k}{dt}\,
+\, \mu_{d \downarrow}^k \frac{dN_{d \downarrow}^k}{dt}
\label{firstPrin}
\end{equation}

For the sake of simplicity,
 we limit our analysis to the isothermal case,
$T^k=T$.  The entropy variation of the sub-layer is deduced from the
two last equations, after introducing the conservation laws and after
defining the polarized currents  $\delta I_{\downarrow } \, = \, (I_{s \downarrow } - 
I_{d \downarrow})/2 $, 
$\delta I_{\downarrow} \, = \, (I_{s \downarrow } - I_{d \downarrow})/2 $, and
the currents $I_{\downarrow} \, =  (I_{s \downarrow } + I_{d 
\downarrow})/2 $,  $I_{s} \, = (I_{s \uparrow } + I_{s 
\downarrow})/2$,

\begin{eqnarray}
T\frac{dS^k}{dt}\,& = &\, \,P_{\phi}^{k-1 \to k}\,-\,P_{\phi}^{k \to 
k+1} \, -\, \frac{1}{2} \Delta \mu_{s}¥ ^k
\left(\delta I^{k-1 \to k}_{\downarrow}- \delta I^{k \to k+1}_{\downarrow}
\,+\, \dot{\Psi}_{sd}¥^k - 2 \, \dot{\Psi}_{s}¥^k \right)
\, \nonumber\\
& & -\, \frac{1}{2} \mu^k_{s} \left (I^{k-1 \to k}_{s}-I^{k \to 
k+1}_{s} - \dot{\Psi}_{sd}¥^k \right )
-\, \frac{1}{2} \Delta \mu_{\downarrow}¥ ^k
\left( \delta I^{k-1 \to k}_{\downarrow}- \delta I^{k \to k+1}_{\downarrow}  - 2 
\dot{\Psi}_{sd}^k - \dot{\Psi}_{s}¥^k ¥\right)
\,  \nonumber\\
& & -\, \frac{1}{2} \mu^k_{\downarrow} \left ( I^{k-1 \to k}_{\downarrow}-I^{k \to 
k+1}_{\downarrow} + \dot{\Psi}_{s}¥^k \right)
\label{entropy}
\end{eqnarray}	

where we have introduce the chemical potentials $\mu_{s}^k\,\equiv
\,\mu_{s \uparrow}^k\, +\, \mu_{s \downarrow}^k$,
$\mu_{\downarrow}^k\,\equiv \,\mu_{s \downarrow}^k\, +\, \mu_{d
\downarrow}^k$, and {\it the chemical affinities of the reactions},
defined by $ \Delta \mu_{s}¥^k\,\equiv \, \mu_{s \uparrow}^k-\mu_{s \downarrow}^k
=- \frac{\partial E^k}{\partial \Psi_{s}¥^k}$, $ \Delta
\mu_{\downarrow}¥^k\,\equiv \, \mu_{s \downarrow}^k-\mu_{d \downarrow}^k =
-\frac{\partial E^k}{\partial \Psi_{sd}¥^k}$. 

The entropy being an extensive variable, the total entropy variation of
 the system is obtained by summation over the layers 1 to $\Omega$ 
 where the layer 1 is in contact to 
the left reservoir $R^l$ and the layer $\Omega$ is in contact to the 
 right reservoir $R^r$. 

The total entropy variation is:

\begin{eqnarray}
T\frac{dS}{dt}\,& = &\, [\ldots]^{R^l \to 1} - [\ldots]^{\Omega \to 
R^r} \nonumber\\ 
& & +\, \sum_{k=2}^{\Omega}\frac{1}{2} \left(\Delta \mu_{s}¥^{k-1}-\Delta 
\mu_{s}¥^k \right ) \, \delta I^{k-1 \to k}_{s} +\, \sum_{k=2}^{\Omega} 
\frac{1}{2} (\mu_{s}^{k-1} - \, 
\mu_{s}^{k})\, I^{k-1 \to k}_{0 s} \nonumber\\
& & + \, \sum_{k=2}^{\Omega}\frac{1}{2} \left(\Delta 
\mu_{\downarrow}¥^{k-1}-\Delta \mu_{\downarrow}¥^k \right ) \, \delta I^{k-1 \to k}_{\downarrow}
 +  \nonumber\\
& &  
\sum_{k=2}^{\Omega} \frac{1}{2} (\mu_{\downarrow}^{k-1} \,- \, \mu_{\downarrow}^{k})\, 
I^{k-1 \to k}_{0 \downarrow} \, + \, \sum_{k=1}^{\Omega} \Delta \mu_{s}¥^{k} \, 
\dot{\Psi}_{s}¥^k \,+ \sum_{k=1}^{\Omega} \Delta \mu_{\downarrow}^{k} \, 
\dot{\Psi}_{sd}^k 
\label{entropytot}
\end{eqnarray}

where the two first terms in the right hand side of the equality 
stand for the heat and chemical transfer from the reservoirs to the 
system $ \Sigma $. 

The entropy variation takes the form
 
\begin{equation}
T\frac{dS}{dt}\, = \,\sum_{i}F_{i} \dot{X}^{i} \,+\, 
P^{ext}(t)
\end{equation}

where $F_{i}$ are generalized forces and $\dot{X}^{i} $ are the conjugated generalized 
fluxes. The variation of entropy is composed by an external entropy
variation $P^{ext}(t)/T$ and by an internal entropy variation 
$dS^{int}/dt$. 

By applying {\it the second law of thermodynamics} $dS^{int}/dt \ge 0$ we
 are introducing the kinetic 
coefficients $L_{\alpha \beta }$ such that 
$dS^{int}/dt=\sum_{ i}F_{i} \left( \sum_{ j} L_{ij}
F^{j} \right)$. By identification with the 
expression~(\ref{entropytot}), the kinetic equations are obtained, 
after performing the continuous limit,

\begin{eqnarray}
\left[\begin{array}{c}
J_{0s} \\
J_{0 \downarrow} \\
\delta J^d_{s}\\
\delta J^d_{\downarrow}\\
\dot{\psi}_{s}¥ \\
\dot{\psi}_{\downarrow} \\
\end{array}\right]
= \left[\begin{array}{cccccc}
L_{ss} & L_{s\downarrow}  & 0  & 0 & 0   & 0\\
L_{\downarrow s} & L_{\downarrow \downarrow}  & 0   & 0  & 0 & 0\\
0  & 0  & L_{\delta s \delta s} & L_{\delta s \delta \downarrow}    & 0 & 0\\
0  & 0  & L^d_{\delta \downarrow \delta s} & L^d_{\delta \downarrow \delta \downarrow}  & 0  & 0\\
0  & 0  & 0 & 0  &  L_{int}^{s}   & 0 \\
0  & 0  & 0 & 0  & 0 & L_{int}^{ \downarrow} \\
\end{array}\right]
\left[\begin{array}{c}
 \frac{-\partial \mu_{s}}{\partial z}\\
 \frac{-\partial \Delta \mu_{s}¥}{\partial z} \\
 \frac{-\partial \mu{\delta }}{\partial z}\\
 \frac{-\partial \Delta \mu_{\downarrow}¥}{\partial z} \\
 {\Delta \mu_{s}} \\
 {\Delta \mu_{\downarrow}} \\
\end{array}\right]
\label{kineticElec}
\end{eqnarray}
 
The kinetic coefficients are state functions; $L_{ij} = L_{ij }(S^k,
N^k_{+},N^k_{-})$ and the symmetrized matrix is positive : $
\frac{1}{2}\,\left\{L_{ji}\,+\,L_{ij} \right\}_{\{ij\}} \ge 0$. The 
coefficients $L_{int}$ refer to the internal relaxation processes 
\cite{DeGroot}. According to Onsager relations, the kinetic coefficients
are symmetric or antisymmetric $L_{ij}=\pm L_{ji}$.  The coefficients 
are known from the two-channel model for the conductivity. The two last
equations concern the internal ($L_{int}$) ``density'' variables 
$\psi_{s}¥ $ and 
$\psi_{sd} $ defined by $\Psi^k = \int_{\Sigma^k} \,
\psi(z)dz $.  Due to the Curie
principle, there is no coupling between spin polarized transport
processes and the electronic transitions (the scalar process is not
coupled to vectorial processes).

\section{appendix B}

This Appendix is devoted to the general resolution of the coupled
diffusion equations Eqs. (\ref{diff}) in the case of a junction between two
semi-infinite layers with the conditions of continuity written below:

\begin{equation}
\begin{array} {lll}
\left ( \frac{\partial \mu_{d \downarrow}¥}{\partial z} \right )_{0^{+}} = 0 \\
J_{s \downarrow}^{N} (0^{-})= J_{s \downarrow }^{F}(0^{+}) \\
\Delta \mu_{s}^{N}(0^{-}) = \Delta \mu_{s}^{F}(0^{+})
\end{array}
\label{condlimite2}
\end{equation} 

Inserting the solutions given by (\ref{soldiff}) in Eqs. (\ref{condlimite2}) and using 
Eqs. (\ref{Onsager1}) the system becomes :

 \begin{equation}
\begin{array} {lll}
b \Lambda_{+} + d \Lambda_{-} = \frac{2 e l_{sd}^{2} 
J}{\sigma_{t} \left ( 1+\beta_{\downarrow} \right )} \\
a L_{s}^{N} l_{sf}^{N} + \frac{\sigma_{s 
\downarrow}¥}{\sigma_{s}}J= -\frac{\sigma_{s \downarrow}¥}{e} 
\left ( -\frac{b}{\Lambda_{+}¥} - \frac{d}{\Lambda_{-}}
\right ) \\ 
b \lambda_{s}^{2} \left ( 
1/l_{sd}^{2} - 1/\Lambda_{+}^{2} \right ) + d \lambda_{s}^{2} \left ( 
1/l_{sd}^{2} - 1/\Lambda_{-}^{2} \right ) = a
\end{array} 
\label{coeflimite}
\end{equation} 

The $b$ coefficient is given by :

 \begin{eqnarray}
b \, \, \Lambda_{+} \left ( \frac{1}{\Lambda_{+}} - \frac{1}{\Lambda_{-}}
\right ) \left [ \frac{1}{l_{sf}^{2}}+\frac{1}{\Lambda_{+} \Lambda_{-}}
+ \frac{\sigma_{s} (1-\beta_{s}^{2}) }{ \sigma_{s}^{N}
(1-(\beta_{s}^{N})^{2}¥) } + \frac{l_{sf}^{N}}{l_{sf}^{2}} \left ( 
 \frac{1}{\Lambda_{+}} + \frac{1}{\Lambda_{-}} \right ) \right ] = 
 \nonumber \\
J_{t} \, \frac{e}{\sigma_{s}^{N}l_{sf}^{2}} \left [  \frac{(1+\beta_{s})}
{(1+\beta_{s}^{N})} l_{sf}^{N}+ \frac{2 \sigma_{s}^{N} l_{sf}^{2} l_{sd}^{2}}
{\sigma_{t} (1+\beta_{\downarrow}) \Lambda_{-} } \left ( 
\frac{1}{l_{sd}^{2}} - \frac{1}{\Lambda^{2}} - \frac{(1-\beta_{s}^{2}) 
\sigma_{s} l_{sf}^{N}}{(1-(\beta_{s}^{N})^{2}) 
\sigma_{s}^{N} l_{sf}^{2} \Lambda_{-}} \right )
\right ]
\label{solcoeflimiteb}
\end{eqnarray} 

the $d$ coefficient :

\begin{equation}
d= J_{t} \, \frac{e}{\sigma_{s}^{N}} \frac{2 
\sigma_{s}^{N}l_{sd}^{2}}{(1+\beta_{\downarrow}) \sigma_{t} 
\Lambda_{-}} - b \, \, \frac{\Lambda_{+}}{\Lambda_{-}}
\label{coeflimited}
\end{equation} 

and the $a$ coefficient :

\begin{equation}
a= - J_{t} \, \frac{2e l_{sf}^{N}}{\sigma_{s}^{N}(1+\beta_{s}^{N})} 
\left ( 1 - \frac{2(1-\beta_{s}) \sigma_{s} 
l_{sd}^{2}}{(1-\beta_{s}^{N})(1+\beta_{\downarrow}) \sigma_{t} 
\Lambda_{-}^{2}}  + 2 b l_{sf}^{N} 
\frac{\sigma_{s}(1-\beta_{s})}{\sigma_{s}^{N}(1+\beta_{s}^{N})}
\Lambda_{+} \left ( \frac{1}{\Lambda_{+}^{2}} -\frac{1}{\Lambda_{-}^{2}} \right )     \right) 
\label{coeflimitae}
\end{equation} 

For $l_{sf} \ll l_{sd}$ :
\begin{equation}
 b \approx \frac{l_{sf}}{2} \frac{eJ_{t}}{\sigma_{s}}
 \frac{(1+\beta_{s})(\beta_{s}-\beta^{N}_{s}¥)}{1
 -\frac{\beta_{s}^{2}+(\beta_{s}^{N})^{2}}{2} }
\end{equation}

For $l_{sf} \gg l_{sd}$

\begin{equation}
 b \approx \frac{eJ}{\sigma_{s}} l_{sd} (1-\beta_{s}^{N}) \frac{1 + 
 \frac{\beta_{s}+ 
 \beta_{s}^{N}}{2}}{1-\frac{\beta_{s}^{2}+(\beta_{s}^{N})^{2}}{2}} 
\end{equation}


\begin{references}
\bibitem{Argu} Eq. (\ref{module}) can also be justified on the 
basis of the Curie principle : the off-diagonal coefficients in the 
Onsager matrix are not zero 
only for variables of identical tensorial nature. In classical 
fluids, the chemical 
affinities (the scalars $\Delta \mu$ in our case) are only coupled 
to the trace of the strain tensor. The 
corresponding term for the magnetization is the modulus of the 
magnetization 
\cite{PRBThermo}.
\end{references}
\end{document}